\documentstyle[preprint,aps]{revtex}
\tighten%
\begin{document}

\title{Faster than light Bell telephone of Michalski transmits only
noise}
\author{{Marek \.Zukowski, Dagomir Kaszlikowski}\\
{\protect\small\sl Instytut Fizyki Teoretycznej i Astrofizyki,}\\
{\protect\small\sl Uniwersytet Gda\'nski, PL-80-952 Gda\'nsk, Poland}\\
}

\date{\today}
\maketitle

\begin{abstract}
Motivated by pedagogical reasons we pinpoint the mistake in the recent
claim, in quant-ph/9911016, that faster than light communication is possible.
\end{abstract}

\vspace{1cm}

In a recent e-print{\footnote{ M. Michalski, quant-ph/9911016}} it is claimed 
that faster than light
communication is possible using EPR-Bell correlated photons. Having in mind
younger students of the field, who may have some difficulties in quick
spotting where  the root of this erroneous claim is, we explicitly pinpoint
it. The other motivation is the fact that exactly this type of mistake 
reappears from time to time in the discussions around the EPR paradox.  
Our reasoning has no originality at all, and 
is just an application of what one can find in many excellent 
textbooks. 

Alice and Bob share an EPR correlated photon pair in the state
\begin{eqnarray}
&&{1\over\sqrt2}(|HH\rangle+|VV\rangle).
\end{eqnarray}
In such a state the probability of Alice to measure a photon
in the state ${1\over\sqrt2}(|H\rangle+|V\rangle)$ is ${1\over 2}$
not $1-\epsilon$ (and this is totally independent of 
what method of measurement is employed). This is because the reduced density matrix of the
photon at Alice side is
\begin{eqnarray}
&&Tr_{Bob}[{1\over\sqrt2}(|HH\rangle+|VV\rangle){1\over\sqrt2}
(\langle HH|+\langle VV|)]=\nonumber\\
&&{1\over2}({1\over\sqrt2}(|H\rangle-|V\rangle))({1\over\sqrt2}
(\langle H|-\langle V|))+{1\over2}({1\over\sqrt2}(|H\rangle+|V\rangle)
({1\over\sqrt2}(\langle H|+\langle V|).
\end{eqnarray}

The above statement gives the final verdict for any attempts
of using EPR correaltions for faster than light communication.
However, in the e-print of Michalski one can find one 
more mistake. The description of the multiport beamsplitter
(fig. 1 of Michalski), in terms of mathematics, is correct. Nevertheless,
the interpretation of the operation of the proposed optical device, 
within the context of quantum measurement is wrong.

The multiport beamsplitter of Michalski in the limit of $n$
going to infinity indeed allows photons which enter it 
(via a chosen port) with 
arbitrary polarization
to exit only either by the $+$ output port of the first beamsplitter, or by
the $+$ output of the last beamsplitter. 
Indeed the polarization state of the exiting photons is the same 
in both cases.
However, in the interpretation of 
these facts Michalski 
completely ignores the following two basic facts.
First,  Michalski  ignores
which of the {\it two} events happens, i.e. behind which 
beamsplitter (the first one or the last one) the photon is registered.
This is a wrong application of the quantum measurement postulates.
The two events are macroscopically distinguishable{\footnote{See e.g.
R.P. Feynman, R.B. Leighton and M.L. Sands, {\it The Feynman Lectures 
on Physics}, (Addison-wesley, Reading, 1963), vol. 3, opening chapter}}!
The second fact is that Michalski forgets altogether that, according to 
his own calculations, {\it which of the two detectors fires is dependent on 
the initial 
state of the photon}. E.g., if $\alpha=\Omega$ the photon
leaves always by the exit port $+$ of the first beamsplitter, and 
if $\alpha=\Omega+\pi/2$ it leaves always via the exit port $+$ of the last 
beamsplitter. In this way the device is indeed measuring (distinguishing) 
the polarization
of the entering photons.
When one forgets, like Michalski, about those facts, and is ignoring the 
information which of the two detectors fired, the device just measures 
the presence of a photon,
but not its polarization. 

When one sets, as Michalski does, $\Omega=\pi/4$, the multiport device 
of Michalski distinguishes perfectly 
between photons in the polarization states 
$\frac{1}{\sqrt{2}}(|H\rangle +|V\rangle)$  and 
$\frac{1}{\sqrt{2}}(|H\rangle -|V\rangle)$. In the first case the detector 
in the $+$ exit port of  the first beamsplitter fires, and in the second 
case the detector behind the $+ $ exit port of the last beamsplitter fires.
I.e., as it was said in the very beginning the probability of Alice to 
measure a photon
in the state ${1\over\sqrt2}(|H\rangle+|V\rangle)$ is ${1\over 2}$
and not $1-\epsilon$.

The fact that on exiting the device the photons are always in the 
same polarization state is completely irrelevant. One could replace 
the complicated
device of Michalski by a single polarizing beamsplitter, no. $1$ in 
the figure of Michalski, and
place behind its $-$ output a suitably orientated half-wave plate, 
which rotates the polarization state by $\pi/2$.  
Two detectors 
should observe the exit ports of this device. And each of them would see only 
photons in
polarization state $\Psi_+$.
Such a simplified device has all the properties of the device of Michalski...,
but in fact is just a plain photon polarization analyzer.

MZ was supported by the University of Gdansk Grant No
BW/5400-5-0264-9. DK was supported by the KBN Grant 2 P03B 096 15.

\end{document}